\documentclass[12pt,epsf]{article}

\usepackage{epsfig}
\usepackage{amssymb}
\usepackage{graphicx}
\usepackage{color}
\usepackage{subfigure}

\begin{document}

\baselineskip 6mm
\renewcommand{\thefootnote}{\fnsymbol{footnote}}


\newcommand{\nc}{\newcommand}
\newcommand{\rnc}{\renewcommand}

\headheight=0truein
\headsep=0truein
\topmargin=0truein
\oddsidemargin=0truein
\evensidemargin=0truein
\textheight=9.5truein
\textwidth=6.5truein

\rnc{\baselinestretch}{1.24}    
\setlength{\jot}{6pt}       
\rnc{\arraystretch}{1.24}   



\newcommand{\tcb}{\textcolor{blue}}
\newcommand{\tcr}{\textcolor{red}}
\newcommand{\tcg}{\textcolor{green}}


\def\be{\begin{equation}}
\def\ee{\end{equation}}
\def\ba{\begin{array}}
\def\ea{\end{array}}
\def\bea{\begin{eqnarray}}
\def\eea{\end{eqnarray}}
\def\nn{\nonumber\\}


\def\ct{\cite}
\def\la{\label}
\def\eq#1{Eq. (\ref{#1})}


\def\a{\alpha}
\def\b{\beta}
\def\g{\gamma}
\def\G{\Gamma}
\def\d{\delta}
\def\D{\Delta}
\def\ep{\epsilon}
\def\e{\eta}
\def\ph{\phi}
\def\Ph{\Phi}
\def\ps{\psi}
\def\Ps{\Psi}
\def\k{\kappa}
\def\l{\lambda}
\def\L{\Lambda}
\def\m{\mu}
\def\n{\nu}
\def\th{\theta}
\def\Th{\Theta}
\def\r{\rho}
\def\s{\sigma}
\def\S{\Sigma}
\def\ta{\tau}
\def\o{\omega}
\def\O{\Omega}
\def\pr{\prime}


\def\half{\frac{1}{2}}

\def\goto{\rightarrow}

\def\na{\nabla}
\def\grad{\nabla}
\def\curl{\nabla\times}
\def\div{\nabla\cdot}
\def\pa{\partial}

\def\ll{\left\langle}
\def\rr{\right\rangle}
\def\lb{\left[}
\def\lc{\left\{}
\def\ls{\left(}
\def\ln{\left.}
\def\rn{\right.}
\def\rb{\right]}
\def\rc{\right\}}
\def\rs{\right)}

\def\vac#1{\mid #1 \rangle}


\def\td#1{\tilde{#1}}
\def\check{ \maltese {\bf Check!}}


\def\Tr{{\rm Tr}\,}
\def\det{{\rm det}}


\def\bc#1{\nnindent {\bf $\bullet$ #1} \\ }
\def\ch {$<Check!>$ }
\def\ss {\vspace{1.5cm}}

\begin{titlepage}

\hfill\parbox{5cm} { }

\vspace{25mm}

\begin{center}
{\Large \bf The dissociation of a heavy meson in the quark medium}

\vskip 1. cm
  {Chanyong Park$^a$\footnote{e-mail : cyong21@sogang.ac.kr}}

\vskip 0.5cm

{\it $^a\,$ Center for Quantum Spacetime (CQUeST), Sogang University, Seoul 121-742, Korea \\}

\end{center}

\thispagestyle{empty}

\vskip2cm


\centerline{\bf ABSTRACT} \vskip 4mm

\vspace{1cm}
We investigate the dissociation of a heavy meson in the medium composed of
light quarks and gluons. In the quark-gluon plasma, the dissociation length of the
heavy meson becomes short as the temperature or quark chemical potential increases.
On the contrary, in the hadronic phase the dissociation length becomes large
as the chemical potential increases, due to the different dissociation mechanism with
one used in the quark-gluon plasma.

\vspace{2cm}


\end{titlepage}

\renewcommand{\thefootnote}{\arabic{footnote}}
\setcounter{footnote}{0}

\tableofcontents

\section{Introduction}

Recently, related to the RHIC and LHC experiments it becomes one of the main
themes to understand the strongly interacting QCD. Many works have been done
using the phenomenological models like the chiral perturbation and
Nambu-Jona-Lasinio model. Another powerful method for this subject
is the lattice QCD based on the numerical manipulation.
At present, using the AdS/CFT correspondence in the string theory \ct{Maldacena:1997re},
various issues related
to the strongly interacting QCD are actively investigated by many researchers
\ct{Sakai:2004cn,Kobayashi:2006sb,Parnachev:2006ev,Karch:2006pv,Kim:2006gp,Nakamura:2006xk,
Horigome:2006xu}.

In the hard wall model, the confinement can be realized by introducing the infrared (IR)
cut-off in the AdS space, which is often called a cut-off AdS \ct{Herzog:2006ra,Domokos:2007kt}.
Using this fact, it was shown that the Hawking-Page
transition of the gravity theory corresponds to the deconfinement
phase transition. The dual geometries are the thermal AdS (tAdS)
for the confining phase and the Schwartzschild AdS black hole (SAdS BH) for
the deconfining phase, respectively. Here, the dual background metrics
do not contain the information for the quarks or baryons
so that the corresponding QCD is a pure gauge theory. On the cut-off AdS or
its deformation,
there were many works to study various physical quantities like the chiral condensation,
isospin matters, etc \ct{Da Rold:2005zs,Csaki:2006ji,Kim:2007qk,Son:2000xc,Son:2000by,Kim:2007gq}.

For the more realistic model, this cut-off AdS space was
improved to more general one including the quark chemical potential or the quark
number density \ct{Kim:2007em,Sin:2007ze,Lee:2009by}. In this case, the backreaction
of bulk gauge field corresponding to the chemical potential of quarks are considered.
The back reacted geometry dual to the quark-gluon plasma becomes the Reissner-Nordstrom AdS
black hole (RNAdS BH). At the low temperature the dual geometry
corresponding to the hadronic phase is the thermal charged
AdS (tcAdS) space, which can be obtained by taking zero mass limit of the RNAdS BH.
In Ref. \ct{Lee:2009by}, the deconfinement phase diagram and the $\r$-meson mass depending
on the chemical potential were investigated. It was also shown that
the critical baryonic chemical potential where the deconfinement phase transition disappears at
zero temperature, is $4000$MeV which is too big to explain the QCD expectation.
In this paper, to cure this mismatch we introduce the slightly different normalization
for the bulk gauge field. Using this, we can find that the critical baryonic chemical
potential becomes $1100$MeV, which is comparable to the QCD result.

In addition, we investigate the dissociation of a heavy meson
affected by the quark chemical potential. As will be shown,
in the quark-gluon plasma
the dissociation length, in which the heavy meson breaks into two heavy quarks,
becomes short as the chemical potential increases.
On the contrary, the dissociation length at the low temperature corresponding to
the hadronic phase, becomes large as the chemical potential increases. Since
there is no free quark in the hadronic phase, the dissociation in this phase implies that
the heavy meson breaks into two bound states of heavy-light quarks. In this procedure,
the pair production of the light quarks is needed so that
the dissociation length increases as the chemical potential becomes large.
These results investigating the finite chemical potential or quark density dependence of
various physical quantities, may be useful and important for the
future experiments like FAIR (Facility for Antiproton and Ion Research).

The rest of paper follows: In the section 2, we shortly review the deconfinement phase
transition in the medium with quarks and gluon. Using the slightly different normalization with
one used in the Ref. \ct{Lee:2009by}, we can show that the baryonic chemical potential
at the critical point, above which there is no deconfinement phase transition, can become
$1100$MeV. This value is comparable to the result obtained from the QCD calculation.
In the section 3, we consider the dissociation of a heavy meson lying in the quark-gluon
plasma using the open string configuration. In this case, the dissociation
length becomes short when the chemical potential or temperature goes up.
In the section 4, we repeat the similar
calculation in the hadronic phase, in which the dissociation length
becomes large unlike the quark-gluon plasma case. We will explain why this behavior can
appears in the hadronic phase.
In the section 5, we finish our work with summarizing the results and
discussing some issues.

\section{Confinement/deconfinement phase transtition in the quark medium}

In the Ref. \ct{Lee:2009by}, it was shown that in the holographic QCD model
the hadronic phase could be described by the thermal charged AdS (tcAdS).
To describe the quark-gluon plasma, we should consider the Reissner-Nordstrom
Ads black hole (RNAdS BH). In this paper, we will investigate the dissociation of
a heavy meson in the medium composed of light quarks and gluons.
Before doing that, we shortly review the deconfinemnet phase transition
with introducing some useful relations
(see the Ref. \ct{Lee:2009by} for the detail).
The Euclidean action describing holographic light quarks is given by.
\be \la{Eact}
S = \int d^5 x \sqrt{G} \lb \frac{1}{2 \k^2} \ls  - {\cal R} + 2 \L \rs
+ \frac{1}{4g^2} F_{MN} F^{MN} \rb ,
\ee
where $G_{MN}$ is given by
\bea    \la{rnbh}
ds^2 &=& \frac{R^2}{z^2} \ls  f(z) dt^2 + d \vec{x}^{ 2}
+ \frac{1}{f(z)} dz^2 \rs  ,
\eea
In the above, the metric function $f(z)$ for the tcAdS is given by
\be
f(z) = 1 + q^2 z^6 ,
\ee
and $f(z)$ for the RNAdS BH is
\be
f(z) = 1 - m z^4 + q^2 z^6 .
\ee
Using this, the Einstein and Maxwell equation are satisfied when
the bulk gauge field $A_t$ is given by
\be \la{solA}
A_{t} \equiv A(z) = i \ls 2 \pi^2 \m - Q z^2 \rs ,
\ee
where $Q$ is related to the black hole charge $q$
\be
Q = \sqrt{\frac{3 g^2 R^2}{2 \k^2}} \ q  .
\ee
In \eq{solA}, it is helpful to introduce the coefficient $2 \pi^2$ for identifying
$\m$ and $Q$ with the chemical potential and the number density
of quark, respectively.

In the case of the fixed chemical potential,
by introducing the Dirichlet boundary conditions at the horizon $A(z_+) = 0$,
$Q$ can be rewritten as
 \ct{Hawking:1995ap}
\be \la{ndch}
Q = \frac{2 \pi^2 \m}{z_+^2} .
\ee
Using this relation, the grand potential of the RNAdS BH becomes
\be
\O_{RN} = - \frac{V_2 R^3}{2 \k^2} \ls \frac{1}{z_+^4} + \frac{8 \pi^4 \k^2}{3 g^2 R^2}
\frac{\m^2}{z_+^2} \rs ,
\ee
where the black hole horizon $z_+$ is given by a function of $T_{RN}$ and $\m$
\be \la{tempzh}
z_+ = \frac{3 g^2 R^2}{8 \pi^4 \k^2 \m^2} \ls \sqrt{\pi^2 T_{RN}^2 +\frac{16 \pi^4 \k^2 \m^2}{3 g^2 R^2}}
- \pi T_{RN}\rs .
\ee
From this, we can find the total quark number density $n_q$
\be \la{tnumd}
n_q \equiv \frac{N}{V_3} = N_c N_f Q .
\ee
Since quark transforms as a fundamental representation under
the U($N_c$) gauge and U($N_f$) flavor symmetry, the above relation \eq{tnumd} is natural.

For the tcAdS case, the boundary condition at the IR cut-off is given by
\be
A (z_{IR}) = - i \pi^2 \m ,
\ee
which relates the black hole charge $q$ to the chemical potential
\be
q =  \frac{3 \pi^2}{z_{IR}^2} \sqrt{\frac{2\k^2}{3g^2 R^2}} \ \m.
\ee
Then, the grand potential for the tcAdS is reduced to
\be
\O_{tc} = - \frac{V_2 R^3}{\k^2} \ls \frac{1}{z_{IR}^4} + \frac{6 \pi^4 \k^2}{g^2 R^2}
\frac{\m^2}{z_{IR}^2} \rs ,
\ee
and the total number density becomes the same form in \eq{tnumd}.

\begin{figure}
\vspace{2cm}
\centerline{\epsfig{file=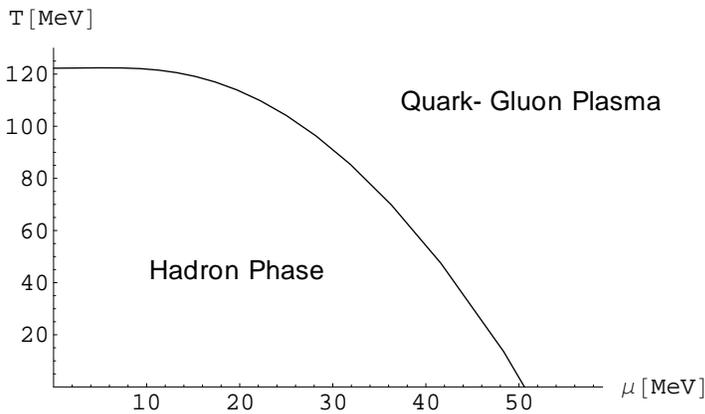,width=12cm}}
\vspace{-4.5cm}
\caption{\small For $N_f/N_c=3/3$, the deconfinement temperature depending on the chemical
potential  }
\label{mesonmass}
\end{figure}

Finally, the chemical potential and the temperature at the deconfinement phase transition
point are given by
\bea
\m_c &=& \frac{1}{2 \pi^2 z_{IR}} \sqrt{\frac{3 N_c}{N_f }
\frac{(z_{IR}^4-2 z_+^4)}{z_+^2 (9 z_+^2 - 2 z_{IR}^2)}} , \\
T_c &=& \frac{1}{\pi z_+} \ls 1 -  \frac{(z_{IR}^4-2 z_+^4)}{z_{IR}^2 (9 z_+^2 - 2 z_{IR}^2)} \rs .
\eea
From the first equation, the range of $z_+$ is given by
$\frac{\sqrt{2}}{3} z_{IR} < z_+ \le \frac{1}{2^{1/4}} z_{IR}$, where $z_{IR} = 1/323$MeV.
For $n \equiv \frac{N_f}{N_c} =1$, we
draw the deconfinement phase diagram in the Figure 1. The critical chemical potential at $T_c=0$ ,
above which there is no deconfinement phase transition,
is $50.8$ MeV. So the total chemical potential including the constituent quark mass
$m_q$
is $\m_{tot} = m_q + \m \approx 390$ MeV and the total baryonic chemical potential becomes
$\m_{B,tot} = m_B + 3 \m \approx 1100$ MeV.
This result is consistent with the QCD expectation.

\section{Heavy meson in the quark-gluon plasma}

From now on, we will investigate the binding energy of a heavy meson in the quark-gluon medium.
In the holographic QCD model, the bound state
of two heavy quarks can be described by the open string ending on the boundary of an
asymptotic AdS space. Since the plasma having light quarks and gluons
can be described by the RNAdS BH in the gravity side, we consider the open
string lying in the RNAdS BH background.

A Nambu-Goto action describing the open string is given by
\be
S = \frac{1}{2 \pi \a'} \int d \ta d \s \sqrt{\det
\pa_{a} X^{M} \pa_{b} X^{N} G_{MN}} ,
\ee
where subindices, $a$ and $b$, are the string world sheet indices and
$G_{MN}$ is the metric of the RNAdS BH background.
We choose the static gauge
\bea
\ta &=& t , \nn
\s &=& x_1=x ,\nn
z &=& z(x) ,
\eea
and assume that the end points of the open string are located at $x = \pm r/2$ in the boundary.
Then, the string action is reduced to
\be
S = \frac{R^2 \b}{2 \pi \a'}   \int_{-r/2}^{r/2} dx \frac{\sqrt{f(z) + z'^2}}{z^2}  ,
\ee
where $\b$ is the periodicity of the time coordinate and
a prime means the derivative with respect to $x$. Notice that for the RNAdS BH background
$f(z)$ is
\be
f(z) = 1 - m z^4 + q^2 z^6 .
\ee
Following the analogy to the particle mechanics,
the Hamiltonian, which is a conserved quantity, is given by
\be \la{ham}
H = - \frac{R^2 \b}{2 \pi \a'} \frac{1}{z^2} \frac{f(z)}{ \sqrt{f(z) + z'^2} } .
\ee
Now, we introduce a parameter $z_0$ which is the maximum value of the open string in the
$z$-direction.
Using the fact that $z'=0$ at $z_0$,
the Hamiltonian at $z=z_0$ becomes
\be \la{hamatmin}
H = - \frac{R^2 \b}{2 \pi \a'} \frac{1}{z_0^2}  \sqrt{f(z_0) }  .
\ee
From \eq{ham} and \eq{hamatmin},
we can easily find a relation between the inter-quark distance $r$ and the maximum
value $z_0$
\be
r = \- 2 \int_0^{z_0} dz \ z^2 \ \frac{\sqrt{f(z_0)}}{\sqrt{f(z)}}
 \frac{1}{\sqrt{f(z) z_0^4 - f(z_0) z^4}} .
\ee
The energy of the heavy meson, $V \equiv S/\b$ is given by
\be
V = \frac{R^2}{\pi \a'} \int_0^{z_0} dz \ \frac{1}{z^2}
\frac{\sqrt{f(z)}}{\sqrt{f(z)  - f(z_0) z^4/z_0^4}} .
\ee
Since this energy diverges at $z=0$, we should renormalize it.

\begin{figure}
\begin{center}
\vspace{2cm}
\hspace{-1cm}
\subfigure{ \includegraphics[angle=0,width=0.5\textwidth]{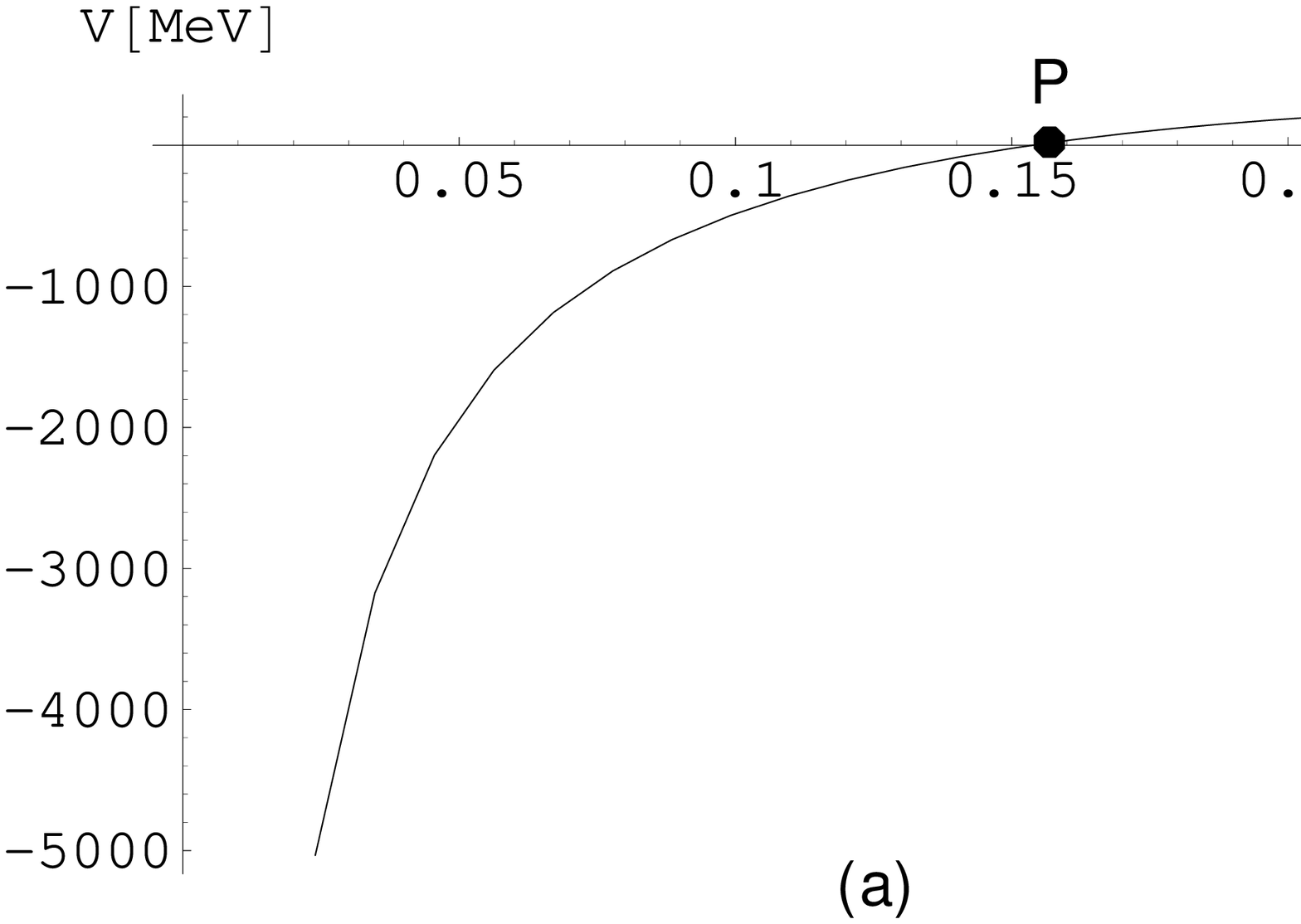}}
\hspace{-0.5cm}
\subfigure{ \includegraphics[angle=0,width=0.5\textwidth]{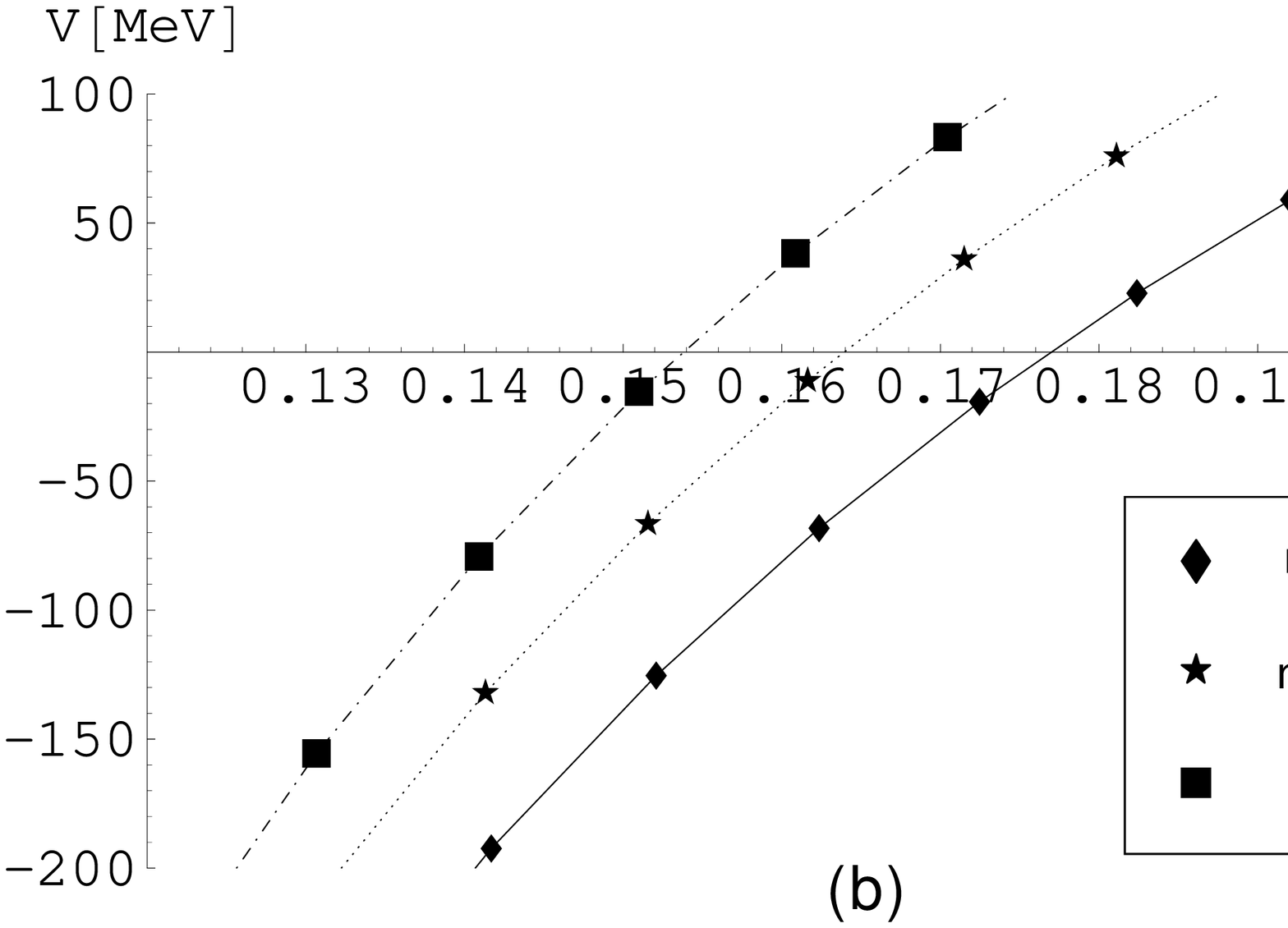}}
\vspace{-2.5cm}
\caption{\small The binding energy depending on the distance between two heavy quarks
in the quark-gluon plasma.
(a) At the point $P$, the dissociation occurs. (b) As the flavor number goes up, the
dissociation length decreases.}
\label{number}
\end{center}
\end{figure}

For this, we consider two straight open strings, which
represent two free heavy quarks and are extended from $z=0$ to $z_+$.
Under the following ansatz,
\bea
\ta &=& t ,\nn
\s  &=& z ,\nn
x_1 &=& {\rm const}
\eea
the energy for two free heavy quarks is given by
\be
V_f =  \frac{R^2}{\pi \a'} \int_0^{z_+} dz \ \frac{1}{z^2} \sqrt{f(z)} ,
\ee
where $z_+$ is the outer horizon of the RNAdS BH.
As a result, the renormalized binding energy of heavy quarks is given by
\bea \la{binden}
V_{b} &=& V - V_f \nn
&=& \frac{R^2}{\pi \a'} \lb \int_0^{z_0} dz \ \frac{1}{z^2}
\frac{\sqrt{f(z)}}{\sqrt{f(z)  - f(z_0) z^4/z_0^4}}
- \int_{0}^{z_+} dz \ \frac{1}{z^2} \sqrt{f(z)} \rb .
\eea
Note that since we consider the static configuration only, the quarks are not
moving in the $x$-direction so that we can ignore the kinetic energy of heavy quarks.
In this paper, since our main interest is to investigate physical quantities
at the dissociation point $V_b=0$, the coefficient $\frac{R^2}{\pi \a'}$
does not play any role.

When the chemical potential and the temperature are given by $\m = 30$MeV and $T=200$MeV,
the binding energy $V_b$ depending on the inter-quark distance $r$ for $n=1$
is shown in the
Figure 2(a), where the point $P$ represents the dissociation of the heavy meson.
In the Figure 2(b), we find that the dissociation length becomes short
as the flavor number $N_f = n N_c$ become large. This implies that
the heavy quarks bound state can be easily dissociated as the flavor number increases.

From now on, we will investigate the dissociation length $r_d$, which is the
inter-quark distance at $V_b=0$, depending on the temperature. For that, we should
note that the binding energy and the inter-quark distance are given as functions
of four variables, $\m$, $T$, $n=N_f/N_c$ and $z_0$.
Therefore, to find the dissociation length depending on
the temperature, we have to fix $\m$ and $n$. Here, we fix $n=1$ and $\m=30$MeV.
In this case, $T$ in the quark-gluon plasma should be higher than $91.2$MeV,
which corresponds to the deconfinement phase transition temperature. From \eq{binden},
we find numerically the set of solutions $\lc z_+, \ z_0\rc$ satisfying $V_b = 0$. Inserting
these solutions into \eq{tempzh}, we can easily read the temperature.
Using these results, we draw the curve representing the dissociation length depending
on the temperature in the Figure 3(a). Notice that since we consider the quark-gluon
plasma, there exists the minimum value of the temperature corresponding to
the deconfinement phase transition temperature. In the Figure 3(a),
the left end of curve represents the deconfinement phase transition point.
As shown in the figure, the dissociation length decreases
as the temperature increases. This is consistent with our intuition. When the temperature
goes up, the heavy quarks bound state should be dissociated easily
due to the thermal fluctuation.

\begin{figure}
\begin{center}
\vspace{2cm}
\hspace{-1.cm}
\subfigure{ \includegraphics[angle=0,width=0.45\textwidth]{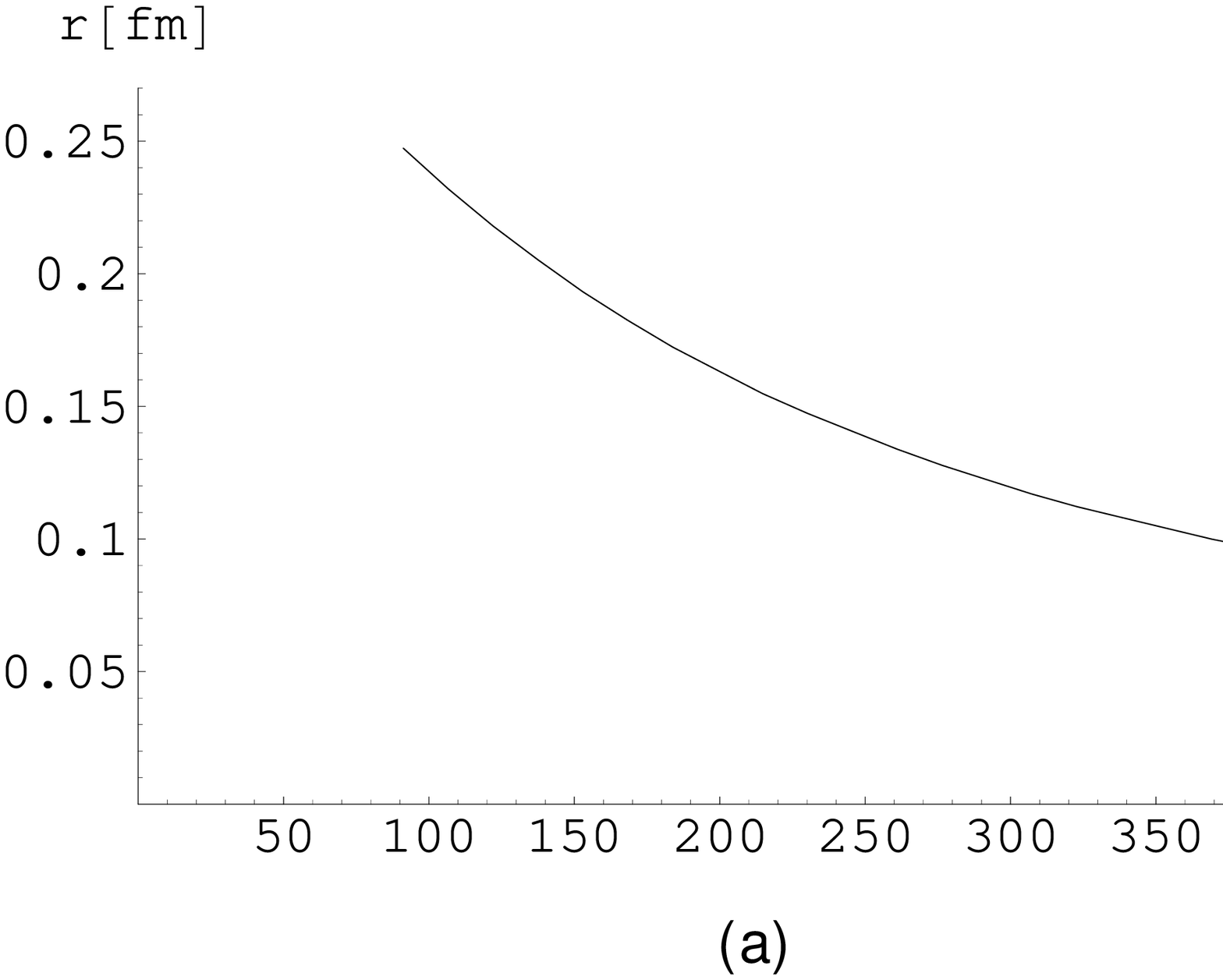}}
\hspace{-0.5cm}
\subfigure{ \includegraphics[angle=0,width=0.45\textwidth]{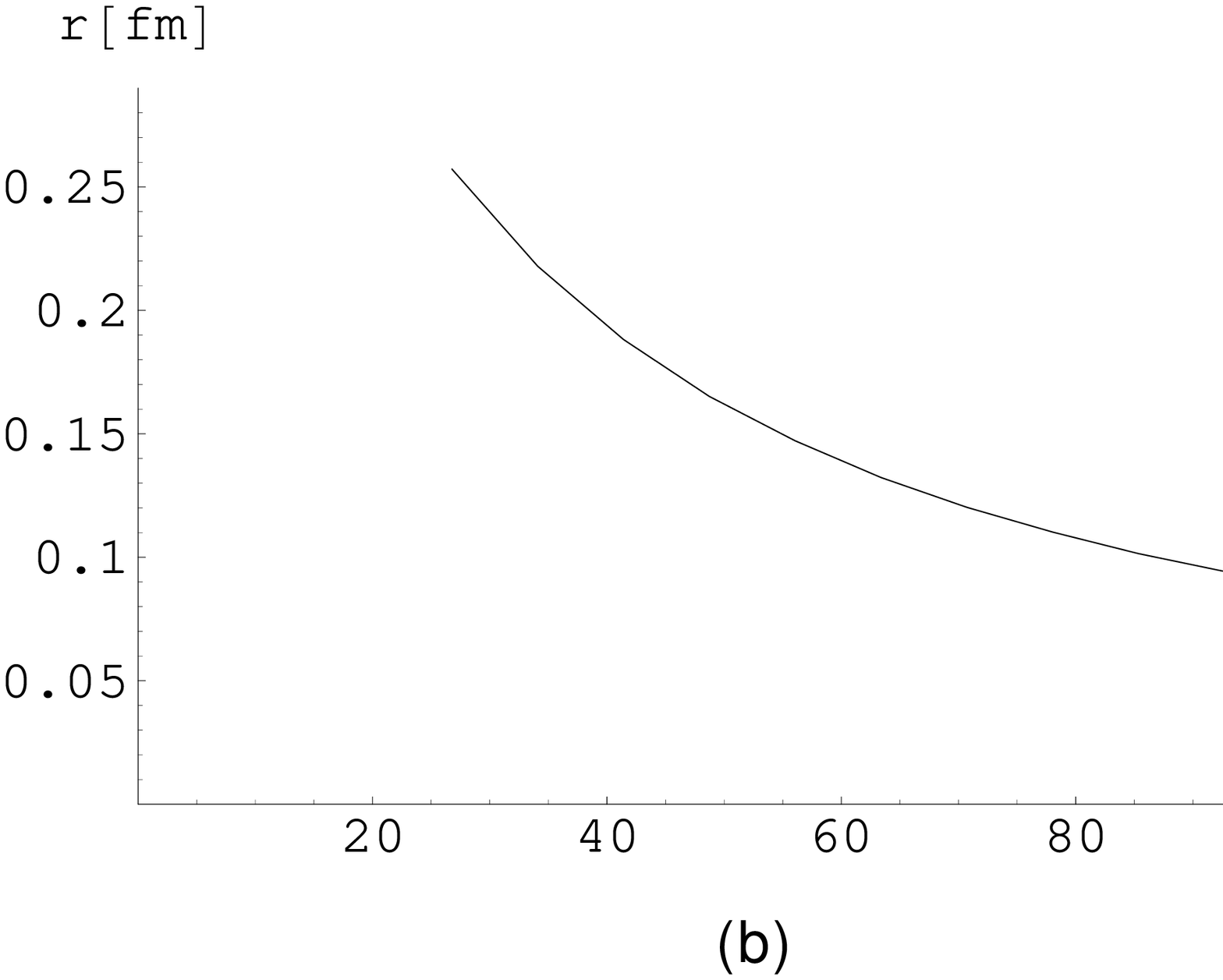}}
\vspace{-2.5cm}
\caption{\small Dissociation length depending on (a) the temperature and (b) the
chemical potential in the quark-gluon plasma.}
\label{number}
\end{center}
\end{figure}

To describe the chemical potential dependence, we fix
$n=1$ and $T=100$MeV. In this case, the deconfinement phase transition occurs
at $\m=26.8$MeV. With the same method in the previous case, the chemical potential dependence
of the dissociation length is shown in the Figure 3(b), where the left end of the plot
is the deconfinement phase transition point. As shown in the figure,
the dissociation length becomes short as the chemical potential increases.
To understand this result, we have to notice from \eq{ndch} and \eq{tempzh}
that the number density is a monotonically
increasing function when the chemical potential increases
\be
Q = \frac{128 \pi^6}{9} \frac{n^2 \m^5}
{\ls \sqrt{\pi^2 T^2 +\frac{16 \pi^4 }{3} n \m^2}
- \pi T\rs^2} .
\ee
Usually, in the deconfining phase the light quarks exist as free quarks, which
disturb the interaction between two heavy quarks. When the chemical
potential becomes large, the number of light free quarks increase and they
screen the interaction between two heavy quarks more strongly. Therefore,
the result in the Figure 3(b) is natural.

\section{Dissociation in the hadronic phase}

\begin{figure}
\begin{center}
\vspace{2cm}
\hspace{-1.5cm}
\subfigure{ \includegraphics[angle=0,width=0.45\textwidth]{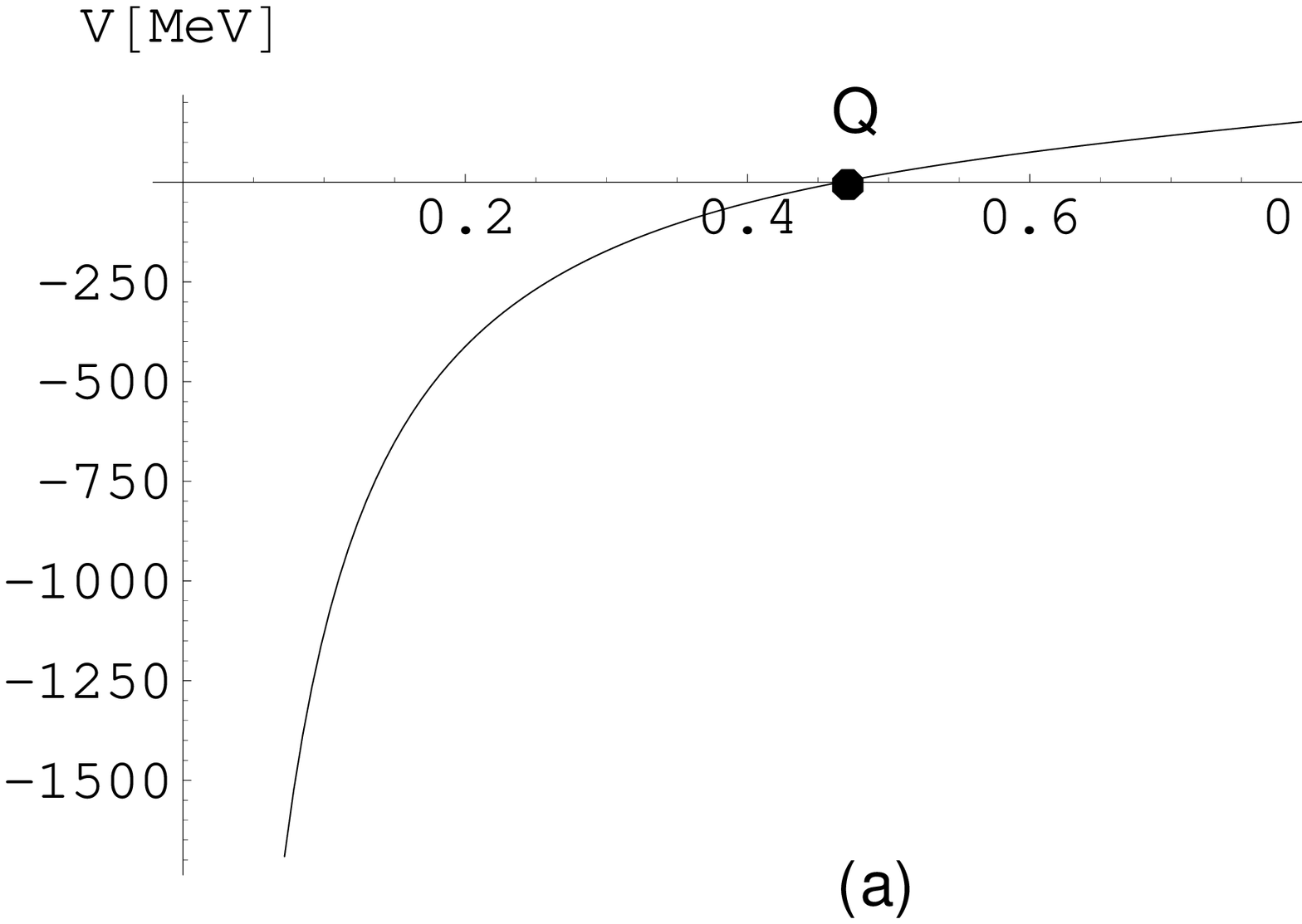}}
\hspace{-0.5cm}
\subfigure{ \includegraphics[angle=0,width=0.45\textwidth]{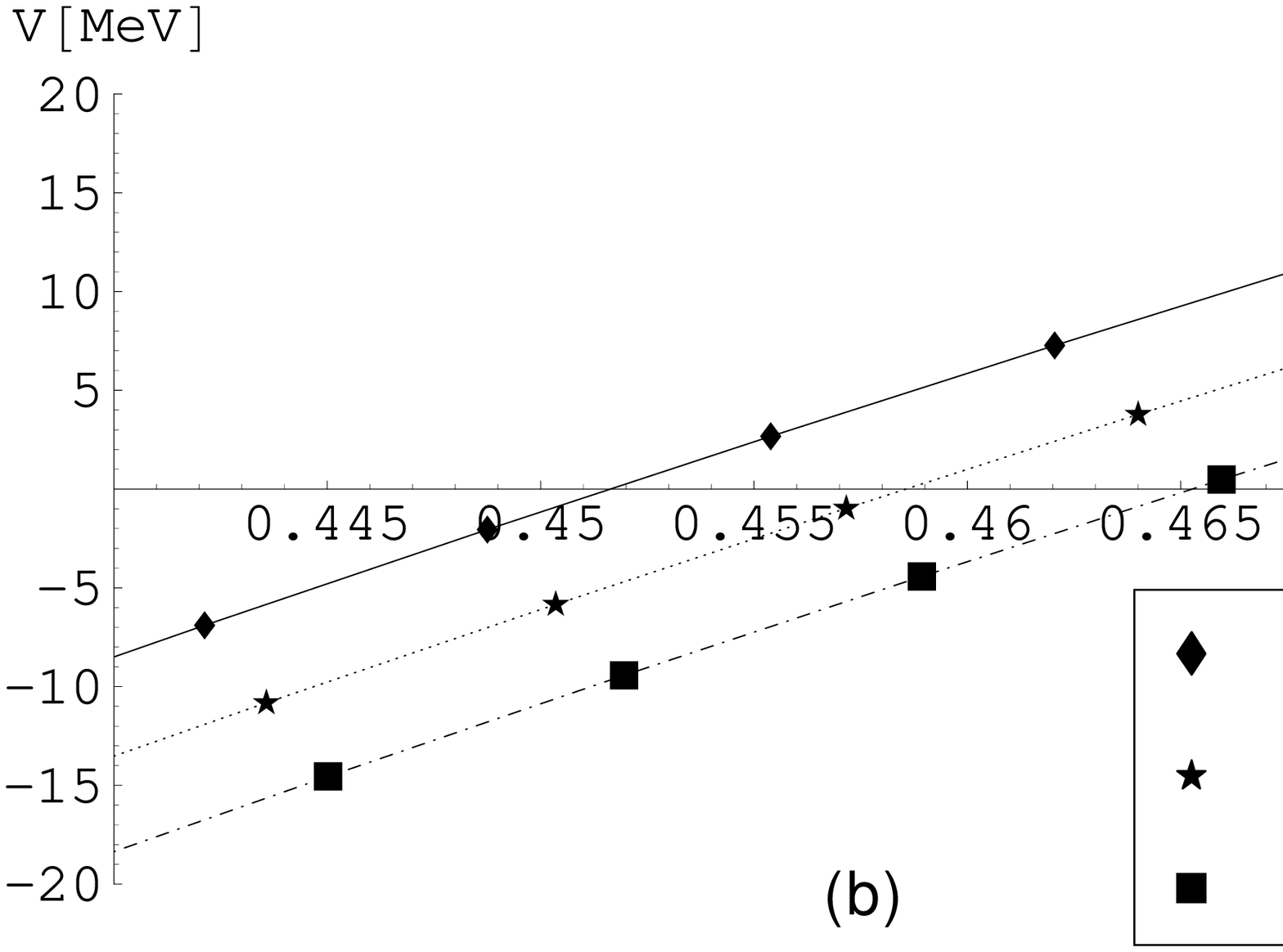}}
\vspace{-2.5cm}
\caption{\small The binding energy of a heavy meson in the hadronic phase. (a)
$Q$ is the dissociation point. (b) It is the binding energy depending on the flavor number,
in which the dissociation length becomes large as the flavor number increases.}
\label{number}
\end{center}
\end{figure}

In this section, we will study the binding energy and the dissociation of a heavy meson
in the hadronic or confining phase. In this case, the dual geometry is the tcAdS.
The binding energy and the inter-quark distance in this background are given
by
\bea
r &=& \- 2 \int_0^{z_0} dz \ z^2 \ \frac{\sqrt{f(z_0)}}{\sqrt{f(z)}}
 \frac{1}{\sqrt{f(z) z_0^4 - f(z_0) z^4}} , \\
V_{b} &=& \frac{R^2}{\pi \a'} \lb \int_0^{z_0} dz \ \frac{1}{z^2}
\frac{\sqrt{f(z)}}{\sqrt{f(z)  - f(z_0) z^4/z_0^4}}
- \int_{0}^{z_{IR}} dz \ \frac{1}{z^2} \sqrt{f(z)} \rb , \la{betcads}
\eea
with
\be
f(z) = 1 + q^2 z^6 .
\ee
Note that the range of $z$ of the second integral in \eq{betcads} is given
by $0 \le z \le z_{IR}$, where $Z_{IR}$ is the IR cut-off.

Using these, we cab find the binding energy depending on the inter-quark distance (see
the Figure 4(a), where $n=1$ and $\m=10$MeV are used. In the hadronic
phase, there is no temperature dependence in the holographic QCD model,
so we consider the zero temperature case only. As shown in the Figure
4(a), there exists a dissociation point $Q$.
Note that there is no free quarks in the confining phase.
Then, what is the meaning of the dissociation in the confining phase? In the deconfining
phase, the dissociation implies that due to the thermal energy
or the screening effect of the light quarks the heavy quarks bound state is broken to
two free heavy quarks. In the confining phase, since there is no free quark, the dissociation
implies that the heavy quarks bound state is broken into two heavy-light quarks
bound state. In the Figure 4(b), we draw the several binding energy
with different flavor number. On the contrary to the quark-gluon plasma case, the
dissociation length increases when the flavor number becomes large. To understand this behavior,
we investigate the dissociation length depending on the chemical potential.
When $n=1$ and $T=0$, the chemical potential dependence of the dissociation length is
shown in the Figure 5. As shown in the figure, the dissociation length increase as
the chemical potential becomes large, which is opposite to the quark-gluon plasma case.
To understand this result, we first notice that the pair creation of two lights quark
is needed to make the heavy meson dissociated.
In the zero temperature the chemical potential can be interpreted as the fermi energy
of the light quark. Therefore, the larger chemical potential becomes
the more energy is needed for the pair creation of light quarks, which make
the dissociation of the heavy meson difficult. So it seems natural that the dissociation length
in the hadronic phase increases as the chemical potential becomes large.
Notice that in the Figure 5. the right end of the plot is the phase transition point.

\begin{figure}
\vspace{2cm}
\centerline{\epsfig{file=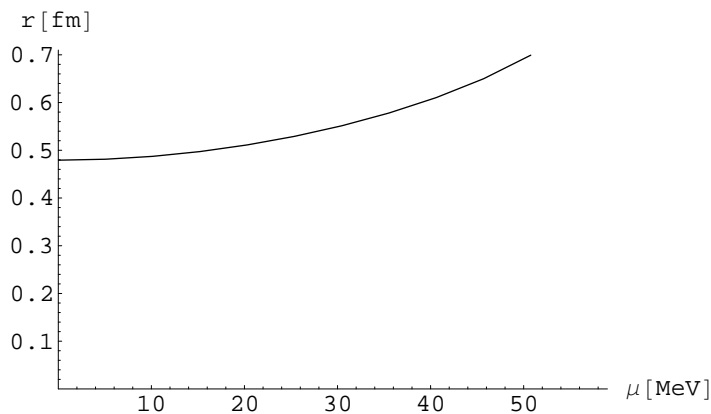,width=12cm}}
\vspace{-4.5cm}
\caption{\small For $n=1$, the chemical potential dependence of the
dissociation length in the hadronic phase.}
\label{mesonmass}
\end{figure}


\section{Discussion}

In this paper, we have shown that the deconfinement phase transition disappears
near the critical baryonic chemical potential $\m=1100$MeV,
which is comparable to the QCD expectation. We also investigated
the dissociation of the heavy meson in the quark-gluon medium. In the quark-gluon plasma,
the interaction between heavy quarks is screened by the light quarks so that
the dissociation length of the heavy meson decreases as the temperature or the light
quark chemical potential becomes large. The dissociation in the hadronic phase implies
that the heavy meson is broken into two heavy-light quark bound states. So the pair creation
of light quarks is needed to dissociate the heavy meson, which makes
the dissociation length become large as the chemical potential increases.

Here, we only considered the backreaction of the quark matters. In general,
there exist other important effects like the chiral condensation and the gluon condensation
in the hadronic phase. To describe more realistic model, we should consider the dual geometry
including these effects and then investigate the various physical quantities.
We hope to report some results of this problem.

\vspace{1cm}

{\bf Acknowledgement}

We would like to appreciate to Youngman Kim, Bum-Hoon Lee and Sang-Jin Sin for valuable
discussion.
This work was supported by the Korea Science and Engineering Foundation
(KOSEF) grant funded by the Korea government(MEST) through the Center for
Quantum Spacetime(CQUeST) of Sogang University with grant number R11-2005-021.

\vspace{1cm}


\end{document}